# Capacitated Kinetic Clustering in Mobile Networks by Optimal Transportation Theory

Chien-Chun Ni, Zhengyu Su, Jie Gao and Xianfeng David Gu

Department of Computer Science, Stony Brook University. {chni, zhsu, jgao, gu}@cs.stonybrook.edu

***Abstract*—We consider the problem of capacitated kinetic clustering in which $n$ mobile terminals and $k$ base stations with respective operating capacities are given. The task is to assign the mobile terminals to the base stations such that the total squared distance from each terminal to its assigned base station is minimized and the capacity constraints are satisfied. This paper focuses on the development of *distributed* and computationally efficient algorithms that adapt to the motion of both terminals and base stations. Suggested by the optimal transportation theory, we exploit the structural property of the optimal solution, which can be represented by a power diagram on the base stations such that the total usage of nodes within each power cell equals the capacity of the corresponding base station. We show by using the kinetic data structure framework the first analytical upper bound on the number of changes in the optimal solution, i.e., its stability. On the algorithm side, using the power diagram formulation we show that the solution can be represented in size proportional to the number of base stations and can be solved by an iterative, local algorithm. In particular, this algorithm can naturally exploit the continuity of motion and has orders of magnitude faster than existing solutions using min-cost matching and linear programming, and thus is able to handle large scale data under mobility.**

## I. Introduction

In this paper, we study the *capacitated kinetic clustering problem* defined as the following: given a set of $k$ base stations with its operating capacity constraints (that may be different), and $n$ terminals, find an assignment of terminals to the base stations such that no base station operates beyond its capacity limit, and that the sum of squared distances between terminals and base stations is minimized.

This is a fundamental problem that has been used to abstract many different application settings in wireless networks. In cellular networks, mobile phones need to connect to static cellular towers or mobile cellular stations; typically a mobile connects to the cell tower or station with the strongest signal strength or the one closest in Euclidean distance. In sensor networks, sensors choose one base station to upload their data. Similarly, often the closest base stations are chosen to minimize energy usage for wireless communication. Here we focus on the three prominent constraints that appear in many of such applications, and our goal is to provide an efficient algorithm for solving this optimization problem in large scale instances.

**Capacity constraints.** Wireless base stations often have fixed capacity constraints which limit the number of users that could be simultaneously served. The limitation may be due to wireless communication limits, such as bandwidth and data rate. Sometimes there may also be constraints imposed by memory limits (bounds on queue size) or power conservation. These physical constraints can differ at different base stations.

For example, the fast growth of smartphones and data usage on cellular networks has shown to be a significant burden on cellular networks since 2009. When the base station (and the network) is overloaded, the result is dropped calls, spotty service, delayed text and voice messages and glacial download speeds. Even with hardware upgrade and the motion from 3G to 4G and 5G, people still observed disrupted or seriously deteriorated service during disasters or emergency situations. In scenarios when the limits set by hardware or protocols are nearly met, we need to carefully schedule and allocate users to these base stations to ensure service quality.

**Energy efficiency.** Our objective for allocation is to minimize the energy consumption by the wireless nodes for wireless communication to the base stations. We follow the free space path loss model in which the signal strength drops in proportion to the square of the distance between transmitter and receiver. For outdoor situation this is a reasonable and convenient approximation to the energy usage of a node communicating to the base station. By using this model, our objective is to minimize the sum of squared distances between the wireless nodes and the base stations they connect to.

**Mobility.** We are particularly interested in supporting mobility of both terminals and possibly the base stations. It has been one of the main challenges for the system configuration to efficiently adapt to node mobility. For example, when pedestrians with mobile devices move around, the mobile loads at terminals change over time. In certain cases, say on new year's eve at Time Square in New York or when at a popular football game, temporal base stations are added to share the traffic load created by the sudden traffic load increase from the terminals. Again, in this case spreading the loads from the moving crowd to mobile base stations becomes a nontrivial problem.

Existing solutions for capacitated clustering are only for the static case. When some of these algorithms are applied for a set of mobile nodes, they will need to be re-computed from scratch, completely missing the continuity and coherence of motion. The sudden change can be disruptive to the service quality. Thus in our setting, we are also interested in not only the optimality of the problem but also the stability of the solution. We would like to explore such a tradeoff in the algorithm design.

This capacitated clustering problem, in fact, is the discrete case of a classical problem studied in the literature, termed the *optimal transportation problem* [1]. In the original problem, one asks for a way of moving a pile of dirt to fill up the holes with the same total volume such that the total cost of transporting this pile of dirt is minimized. The optimal solution for this problem defines a mapping from an input domain (i.e., the distribution of dirt or generally a probability measure) to an output domain (i.e., the distribution of holes). The transportation cost, in Brenier's formulation [2], is precisely the quadratic Euclidean distance between a point of the input domain to its mapped position. Therefore we can solve our allocation problem by using algorithms for solving the optimal transportation problem.

There are already algorithms developed in the literature for the optimal transportation problem especially in the discrete setting when weighted terminals are assigned to stations of fixed capacity. In a simple case, when the weights are all 1 and capacities are integer numbers, we can define a complete bipartite graph on the mobile terminals and duplicate a base station $j$ to $cap(j)$ copies, where $cap(j)$ is the capacity of $j$. The edge between terminal $i$ and base station $j$ is the squared Euclidean distance. In this case the optimal transportation problem is precisely the min cost matching problem and can be solved in $O(n^3)$ time, where $n$ is the total number of vertices of the bipartite graph. In general the problem can be defined and solved by linear programming, which is the state of art solution.

The LP solver for capacitated clustering might be alright for the static, one-shot scenario, especially when one uses a highly optimized LP solver. But such an algorithm will be inappropriate for the kinetic setting when the terminals or base stations move around. There is no obvious way to exploit the continuity of motion in the LP algorithm and one has to recompute from scratch. When the number of terminals is large, the running time could be potentially a huge burden.

The main contribution of this paper is to make use of the structural insights in the optimal transportation theory, which leads to a fast algorithm for the capacitated clustering problem. The algorithm turns out to be orders of magnitude faster than the algorithm using LP solver. Due to the geometric nature of our algorithm, it is extremely well suited for the kinetic setting. To get an idea, we start from the case when the capacity constraints are lifted. In this case, clearly the objective is minimized when each mobile terminal is allocated to its closest base station. That is, the allocation is obtained by the Voronoi diagram on the base stations. This means that the solution can be represented in size $O(k)$. Here $k$ is the number of base stations, and is usually much smaller than $n$, the number of mobile terminals under consideration. As it turns out, when the capacity constraints are imposed, the allocation is still represented by a Voronoi-type convex decomposition – the weighted Voronoi diagram or termed power diagram. In a power diagram, the sites are weighted either positively or negatively. The power distance from a point $p$ in the plane to a site $q$ with weight $h_i$ is $|pq|^2 - h_i$. The power diagram is the partitioning of the plane such that each point is grouped to its closest site by the power distance. Therefore the optimal allocation is again represented by size $O(k)$.

By using this formulation of power diagram, we can study optimal capacitated clustering in the mobile setting. We develop an iterative, *distributed* algorithm to solve for the target power diagram. The only parameter that each base station $i$ would need to adjust is the weight $h_i$. If the current power cell has too many terminals that what this base station can handle, base station $i$ would decrease its weight, which increases the power distance and thus force some of the terminals to be assigned elsewhere. In previous work, there were two approaches to solve this optimization problem: the first one is a centralized algorithm that uses Newton's method [3]; it is very fast. The second one uses gradient descend method [4], it is naturally distributed but slower. In this paper we elaborate a new approach implementing the Newton's method. It is distributed *and* very fast.

When nodes move around, we exploit the continuity of motion in the sense that the allocation solution for the $i$th snapshot can use the optimal solution at the $(i-1)$th snapshot as the initial value used in the optimization procedure. This greatly reduces the number of iterations in finding the optimal solution. In contrast, both LP and min cost matching have no easy way of utilizing previous solutions, and have to restart from scratch. Using kinetic data structure we also provide upper bounds on the number of changes of the optimal solution when the terminals and base stations move along pseudo-algebraic trajectories.

## II. Capacitated Kinetic Clustering Problem

Consider a set $X$ of $n$ wireless nodes/terminals $X = \{x_1, x_2, ..., x_n\}$ and a set $Y$ of $k$ base stations $Y = \{y_1, y_2, ..., y_k\}$. Each base station $y_j$ has a fixed capacity $cap(y_j)$, limiting the number of terminals it can serve. All capacities sum up to be the total number of terminals: $\sum_j cap(y_j) = n$. We would like to assign the terminals to the base stations such that all capacity constraints are satisfied (no base station is overloaded) and that the sum of squared distance is minimized. In particular, we look for a clustering of nodes into $k$ clusters, $X_1, X_2, \cdots, X_k$. Nodes in $X_i$ are assigned to base station $y_i$. This clustering minimizes the sum of squared distances from each terminal to its assigned base station.

$$\begin{array}{ll} \min & \sum_j \sum_{x_i \in X_j} ||x_i - y_j||^2 \\ \text{s.t.,} & |X_j| = cap(y_j), \quad \forall 1 \le j \le k \end{array}$$

In this paper, we focus on the setting when both base stations and the terminals can possibly move around. We would like to maintain the optimal transport solution at all times. This is termed the *capacitated kinetic clustering* problem.

In this paper we assume that $n \gg k$. We assume that the base stations are connected using out band channels such that they can collaboratively compute the optimal clustering solution and inform the terminals.

In the following we first review the connection of the (static) capacitated clustering problem to optimal transport theory. Then we discuss our contribution, which has two parts. Section IV describes our distributed algorithm for kinetic capacitated clustering problem and we show its efficiency in the simulation section. Section VI describes the lower bound and upper bound on the number of changes to the kinetic problem.

## III. Optimal Transport Theory

The capacitated clustering problem is a special case of the general optimal transport problem, originally proposed by Monge back in 1781 [1], [5] and later revised by Kantorovich [6]. This is termed Monge-Kantorovich problem. Given two domains $X$ and $Y$ with the corresponding density measure $\mu$ and $\nu$, respectively, the transportation cost from $x \in X$ to $y \in Y$ is defined as $cost(x, y)$. The optimal transport plan to this problem is a measure $\gamma$ on $X$, $Y$, such that $\forall A \subset X$ and $\forall B \subset Y$, $\gamma(A \times Y) = \mu(A)$, $\gamma(X \times B) = \nu(B)$, where $\gamma(A \times B)$ represents the partial (or total) masses to transport from $A$ to $B$. The total transportation cost to minimize is

$$\int_{X \times Y} cost(x, y)\mu(x)d\gamma(x, y).$$

For this problem Kantorovich proved the uniqueness and existence of the optimal solution.

For the discrete version of the Monge-Kantoroich problem, i.e., $X = \{x_1, x_2, ..., x_n\}$ and $Y = \{y_1, y_2, ..., y_k\}$, with the Dirac measures $\mu = \{\mu_i\}$, $\nu = \{\nu_j\}$ and total mass of 1 each, an optimal transportation plan is represented by a $n \times k$ allocation matrix $\gamma = [\gamma_{ij}]$, where $\sum_i \gamma_{ij} = 1$, $\sum_j \gamma_{ij} = 1$. The element $\gamma_{ij}$ in $\gamma$ represent $x_i$ with the percentage of mass $\mu_i(x_i)$ delivered to the location $y_j$, and the discretized total transportation cost is $\sum_{ij} cost(x_i, y_j)$. This problem can be solved by linear programming method with $n \times k$ variables.

In late 1980's, Brenier proposed a novel geometric approach for the Monge-Kantorovich problem by using the geometric characteristics of the domain [2]. When the cost function $cost(x, y)$ is the quadratic Euclidean distance $cost(x, y) = ||x - y||^2$, there exists a convex function $f : X \to \mathbb{R}$ such that the optimal transport map is given by the function's gradient map $x \to \nabla f(x)$. Furthermore, the optimal mass transportation map is unique. By Brenier's theorem, the Monge's problem is converted to solving the following Monge-Amperé partial differential equation:

$$\mu(x) \det \left( \frac{\partial^2 f(x)}{\partial x_i \partial x_j} \right) = \nu \circ \nabla f(x).$$

In the discrete setting (as shown above), the convex function as described in Brenier's theorem is in fact closely connected to the power diagram in the Euclidean plane. To see that, consider the special case of our capacitated clustering problem when the capacity constraint is lifted away. The optimal solution to minimize the transport cost is obvious – send each terminal to its closest base station. This solution is exactly defined by the Voronoi diagram – partitioning of the Euclidean plane into convex cells such that the points of each cell have the same *closest* base station in $Y$. When the base stations have capacity constraints, the solution is in fact defined by power diagram, i.e., Voronoi diagram when the sites have weights. Here each base station $y_j$ has a weight $h_j$. And the *power distance* from each terminal $x_i$ to $y_j$ is defined by $\text{Pow}(x_i, y_j) = ||x_i - y_j||^2 - h_j$. The resulting partition of the plane such that each cell has the same closest base station when power distance is used is called the *power diagram*. It is known that the power diagram is still a convex partitioning and each cell is shrinking/expanding when the weight of base station is decreased/increased. The optimal transport solution in fact is precisely defined by the power diagram of certain weights, such that the total number of terminals in the cell for $y_j$ is precisely the capacity $cap(y_j)$.

Voronoi diagram and power diagram have a nice lifting map that shows the direct connection to Brenier's theorem. Let $f : z = ||x||^2/2$ as the paraboloid in $3D$. Now project each base station $y_j$ to the point $y'_j$ on the paraboloid, with $z$-coordinate $||y_j||^2/2$. Now we take the tangent plane of the paraboloid at $y'_j$. Then the upper envelope of this arrangement of planes, when projected down to 2D, is the Voronoi diagram. For power diagram with weight $h_j$'s, the tangent plane corresponding to $y_j$ is shifted down by amount $h_j$. The projection of the upper envelope is precisely the power diagram. Recall that Brenier's theorem characterized the solution to the optimal transport problem is given by the gradient map $x \to \nabla f(x)$ of a convex function $f : X \to \mathbb{R}$. In fact, this convex function $f$ is the upper envelope of the lifted hyperplanes for the power diagram, and the optimal transport solution is specified by the power diagram.

The existence and uniqueness of the solution was proven multiple times using different approaches by Alexandrov [7], Armstrong [8], Brenier [2]. The connection to power diagram was first made by Aurenhammer et al. [4]. Recently, the existence and uniqueness were proved by Gu et al. [3] using variational principle. However, none of the algorithms derived from the proofs are distributed.

## IV. Distributed Algorithm

In the following we provide the first distributed algorithm for the capacitated kinetic clustering problem, or, a discretized version of Monge-Kantorovich problem. Let's first discuss the centralized setting and the algorithm by using variational principle, first reported in [9]. In the next subsection we describe the distributed algorithm and the proof for correctness.

### A. Centralized Algorithm using Variational Principle

Suppose that the capacity of base station $i$ is $\bar{A}_i$. To find the clustering scheme for base stations $Y$ with terminals $X$ on domain $D$, by Monge-Briener theory, we try to find the height vector $\mathbf{h} = (h_1, h_2, ..., h_k)$ where the supporting planes for each base station $y_j$ are

$$\pi_j(\mathbf{h}) :< x, y_j > +h_j.$$

We take the upper envelope of all such planes in $\mathbb{R}^3$, which is a convex function denoted as

$$u_{\mathbf{h}}(x) = \max_i < x, y_j > +h_j.$$

This convex function induces a polygonal partitioning of $D$: $D = \cup_j W_j(\mathbf{h})$.

The polygonal partition $W_j(\mathbf{h})$ introduces convex cells. All the terminals in the same cell projected by $\pi_j(\mathbf{h})$ are assigned to the corresponding base station $y_j$. We define the area of this cell as the number of terminals inside and form the gradient function:

$$\nabla E(\mathbf{h}) = (Area(W_j(\mathbf{h}) \cap D)).$$

Now we compute the Hessian matrix $H(\mathbf{h}) = h_{ij}(\mathbf{h})$. For that we compute the cell's neighboring edge length $e_{ij} = W_i \cap W_j \cap D \neq \emptyset$, and the corresponding dual triangulation vertices distance $\overline{e_{ij}}$. Define $w_{ij}$ as $w_{ij} = |e_{ij}|/|\overline{e_{ij}}|$ for $i \neq j$, $W_i \cap W_j \cap D \neq \emptyset$. Now we have by variational principle

$$h_{ij}(\mathbf{h}) = \begin{cases} -w_{ij} & i \neq j,\, W_i \cap W_j \cap D \neq \emptyset \\ \sum_\ell w_{ik\ell} & i = j \\ 0 & otherwise \end{cases}$$

Last, we iteratively apply Newton's method to update the height vector $\mathbf{h}$:

$$\mathbf{h} \leftarrow \mathbf{h} + \varepsilon H(\mathbf{h})^{-1} \nabla E(\mathbf{h}),$$

where $\epsilon$ is the step length of Newton's method. The update process stops and outputs the final result until the mean square error of $(w, \overline{w})$ converges to $\delta w$.

Notice that one can use a centralized optimization method to compute the Newton's step and update the height vector $h$ iteratively. The correctness of this algorithm is proved in [3]. In the next section we show a distributed algorithm in which each base station adjusts its own height value, which essentially also implements the Newton's method.

### B. Distributed Newton's Method

Instead of a centralized view to compute the height vector $\mathbf{h}$ of all base stations, in the distributed algorithm, each base station repeatedly adjusts its individual height (the weight for power diagram) to find the power cell to fit the clustering requirement.

The optimal $\mathbf{h}$ we are looking for is the unique global optimum of the following convex energy:

$$E(\mathbf{h}) = \int^{\mathbf{h}} \sum_{j=1}^{k} (\bar{A}_j - A_j(\mathbf{h})) dh_j,$$

where $A_j(\mathbf{h})$ is the area of the cell $W_j(\mathbf{h})$, $\bar{A}_j$ is the desired area/capacity of the cell.

Rewrite the Newton's method. We have $\nabla E(\mathbf{h}) = H\vec{x}$, where $\vec{x} = \delta\mathbf{h}$ for step size $\delta$. On the left hand side:

$$\nabla E(\mathbf{h}) = (\bar{A}_1 - A_1, \bar{A}_2 - A_2, \cdots, \bar{A}_k - A_k)^T.$$

Consider only element at position $i$, we have

$$\bar{A}_i - A_i = \sum_j w_{ij}(x_j - x_i).$$

Reorganize, we have

$$x_i = \frac{\sum_j w_{ij} x_j - (\bar{A}_i - A_i)}{\sum_j w_{ij}}.$$

Thus we would like to obtain the value $x_i$ locally. To do that we use a local linear iterative method. Essentially,

$$x_i^{(n+1)} = \frac{\sum_j w_{ij} x_j^{(n)} - (\bar{A}_i - A_i)}{\sum_j w_{ij}}.$$

This iterative method eventually converges to $x_i$.

Once each base station $i$ has its own update vector $x_i$, they can update their height value, and one single step of Newton's method is then accomplished. In this way the Newton's method can be completely solved by local, iterative, distributed methods.

We remark that as the Hessian matrix $h_{ij}(\mathbf{h})$ is positive definite, the energy is convex in the space

$$\{\mathbf{h} | \sum_{j=1}^{k} h_j = 0, A_j(\mathbf{h}) \geq 0, \forall j\}.$$

Therefore one can also use the gradient descend method

$$\mathbf{h} \leftarrow \mathbf{h} + \epsilon \nabla E(\mathbf{h}) = \mathbf{h} + \epsilon(\bar{A}_1 - A_1(\mathbf{h}), \cdots, \bar{A}_k - A_k(\mathbf{h}))^T$$

to compute the global optimum. This gradient descend method turns out to be the same as the iterative method proposed in [4] although the two methods employ different energy functions. Further, using the distributed Newton's method is much faster than the gradient descend method. As can be shown in the evaluation section, the speedup (in terms of the number of iterations) can be one or two orders of magnitude.

### C. Iterative Algorithm for Mobile Terminals

To acquire the clustering scheme for mobile terminals, we sample the movement of terminals into a series of snapshots, and process them iteratively. Suppose terminals move continuously, the distribution of terminals between two contiguous snapshots should be similar. Thus, the power diagrams between these two snapshots are roughly the same. We can apply the final power vector $\mathbf{h}$ from previous snapshot for a better initial state while computing the power diagram of one new snapshot to iteratively reuse previous result.

We can also assign a capacity tolerance value for base station as a tradeoff with the computational time. That is, by adjusting the exactness of clustering, after acquiring the initial scheme, the iterative process only takes action while the base station is over the capacity of a given tolerance percentage.

## V. Evaluation

In this section we evaluate our optimal transport clustering scheme (Optran) result along with the LP, the perfect matching and gradient descend solution. We consider both static and dynamic case of terminals under different shapes of domains, and compare the computation time and the final distance squared sum as energy consumption between each method. Our observations are summarized as follows:

1) For the capacitated kinetic clustering problem, Optran is up to 10000 faster than LP and perfect matching solution. That is, only 0.08 seconds to allocate 8000 terminals to 8 base stations.
2) While in LP and perfect matching, the computation time grows polynomially with the terminal size, Optran grows only linearly, and provides energy consumption as good as other two solutions.
3) In mobility case, by reusing previously acquired power vector, up to one half of computation time for a series of snapshot can be reduced.
4) By applying proper capacity tolerance for each base station, the total computation time for mobile case can be reduced up to 30%.

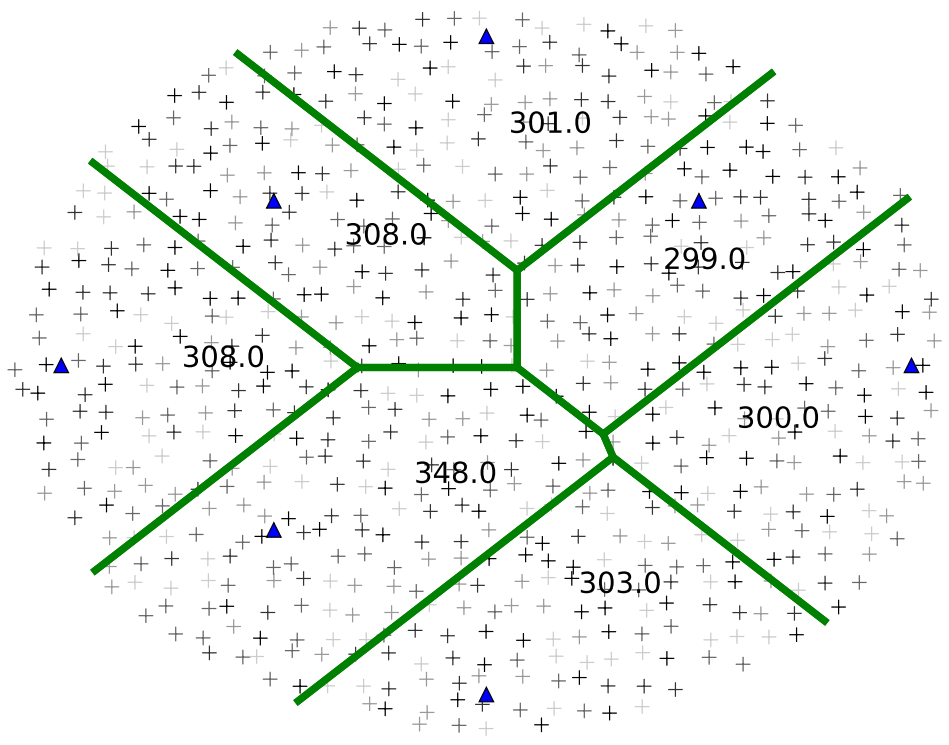

(a) A standard disk domain, base stations are assigned with different capacity, terminal with different data usage are label with colors from light to dark. The terminals are distributed as perturb grid.

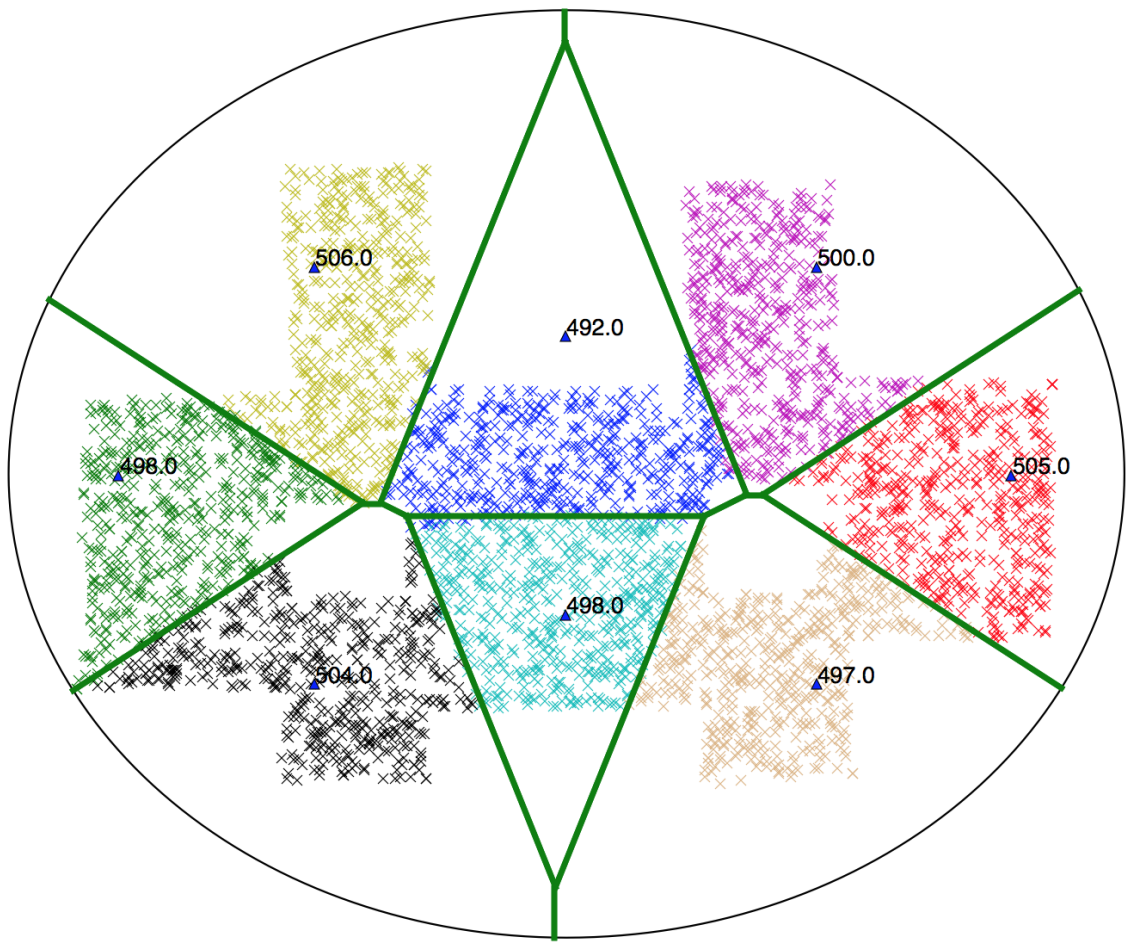

(b) A simple polygon domain with holes, rescaled to the disk domain. The terminals are randomly distributed in to the domain. The requested capacities of each base station are assigned to be equal in this case. The result from LP is shown as the difference of terminal colors.

Fig. 1. Examples for capacitated kinetic clustering with Optran. The cell surrounding by the green lines represent the covering area for the contained base station marked with blue triangle. The base station is labeled with the actual terminals assigned to it.

### A. Simulation Setting

Our clustering scheme Optran is performed on C++, supported by the *CGAL* [10] library. The comparison results of LP are acquired by the industry standard LP solver *CPLEX 12.3* [11], *Gurobi 5.5* [12], and the MATLAB built-in LP solver, both operating in *MATLAB R2013a* [13]. Since the results of all of the solver are similar in our experience, we represent the LP result only by the $CPLEX$ solution. For perfect matching solutions, the results are acquired by the built-in Hungarian algorithm of MATLAB. All program ran on Retina Macbook Pro 2012 with 2.6 GHz Intel Core i7 and 16GB RAM, operating on OSX 10.8.4.

We follow the free space path loss model in a free space. In this model, the signal strength is proportional to the square of the distance between transmitter and receiver. Since signal strength is also proportional to the energy cost, in here, we use the *distance squared sum* to represent the energy cost. To simulate the open space domain, terminals – even with different data usages, are spread randomly or based on perturb grid in domain. More, we arbitrarily distribute the base stations with different capacity requirement in given domain. An example of domain is illustrated fig. 1. Since our method is suitable for any simple domain even with holes, the only requirement to run Optran is to shrink the input domain into an unit disk.

| Iteration | Mean Squared Errors | Energy Cost | Time (sec) |
|---|---|---|---|
| 1 | 0.129422 | 255.277 | 0.005652 |
| 2 | 0.0217536 | 224.503 | 0.004595 |
| 3 | 0.00505614 | 216.775 | 0.004610 |
| 4 | 0.0010133 | 212.869 | 0.004592 |
| 5 | 0.000345268 | 212.153 | 0.004587 |
| 6 | 0.000224121 | 211.735 | 0.004522 |
| 7 | 0.0000199027 | 211.619 | 0.004552 |
| Sum : | | | 0.033110 |

TABLE I. DETAILED ITERATIVE PROCESS, $\delta w = 1e - 4$.

For observation and comparison conveniences, in the following experiment we only consider disk domain with equal base station capacity requirement, and the $\delta w$ is set to be $1e^{-4}$.

*1) Linear Programming solution:* For linear programming (LP) solution setting, we follow the discretization process of Monge-Kantorovich approach mentioned before.

Given weighted terminals $X = \{x_1, x_2, ...x_n\}$ and capacitated base stations $Y = \{y_1, y_2, ...y_k\}$, we assume the transportation plan to be a $n \times k$ matrix $\rho_{ij} \geq 0$. $\rho_{ij}$ acquires the following LP equations:

$$\begin{aligned}
\text{Minimized:} \quad & \sum_j (\sum_i dist(x_i, y_j) \rho_{ij} x_i) \\
\text{Subject to:} \quad & \sum_i \rho_{ij} cap(x_i) = cap(y_j) \\
& \sum_i \rho_{ij} = 1 \\
\text{Bounds:} \quad & \forall i, j \quad 0 \leq \rho_{ij} \leq 1
\end{aligned}$$

*2) Perfect Matching solution:* We apply the famous Hungarian algorithm [14] for minimized weight bipartite perfect matching solution. To construct the proper bipartite graph for the matching $M : X \to Y$ such that $|X| = |Y|$, for a base station $y_j$ with the capacity $cap(y_j)$, we duplicate the base station in the graph $cap(y_j)$ times. The weight of the graph is assigned to be the distance $dist(x_i, y_j)$, which is the distance from the terminal $x_i$ to base station $y_j$. Notice that since the data from one terminal are not splittable, we apply the perfect matching for the case that each terminal only have 1 data.

### B. Simulation Result

*1) Power Vector Iteration:* The computational cost for Optran is dominated by the iterative process of acquiring the required power vector for power diagrams. With the proposed energy function along with the gradient decent method, we are able to retrieve the optimal result within constant iterations. Table I list the detailed iterative process of computing the clustering scheme for fig. 9(a), the $\delta w$ is set to be $1e - 4$, 7 runs are required. While the mean squared errors $\delta w$ is given, the iterations stop only when $\delta w$ requirement is met. From iteration to iteration, the energy cost is monotonically decrease to an optimal value.

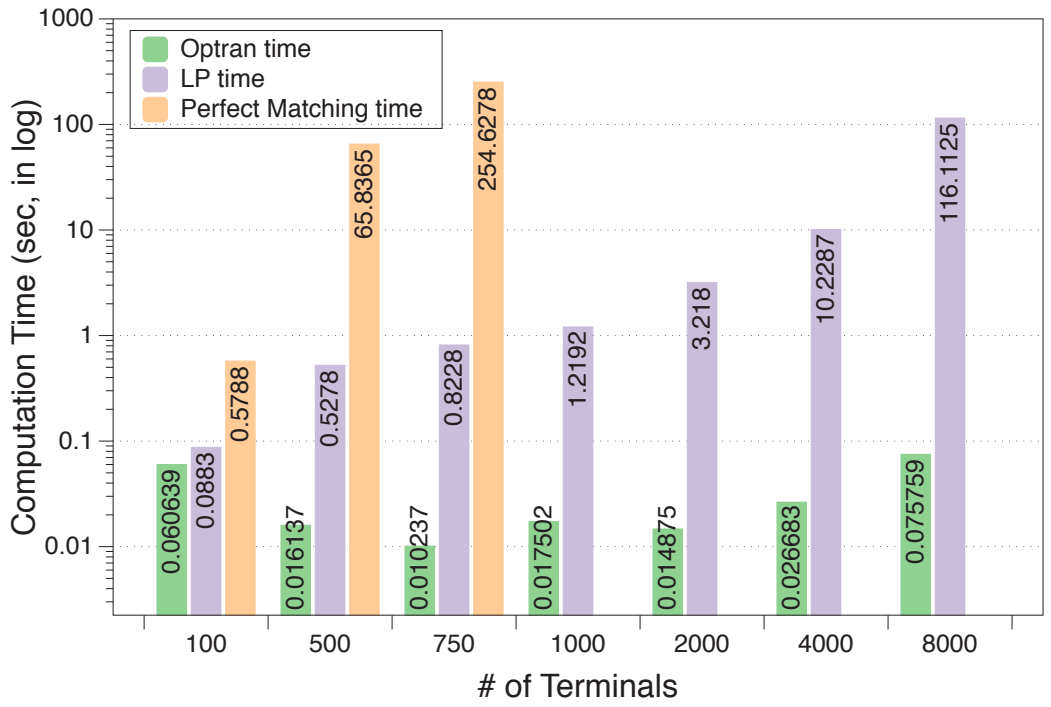

Fig. 2. A comparison of computation time between Optran, LP, and perfect matching on different sizes of terminals. Number of base stations is fixed at 8, with equal capacity constraints. Notice that for both LP and perfect matching solution, the time growths exponentially compare with the linear growth with Optran.

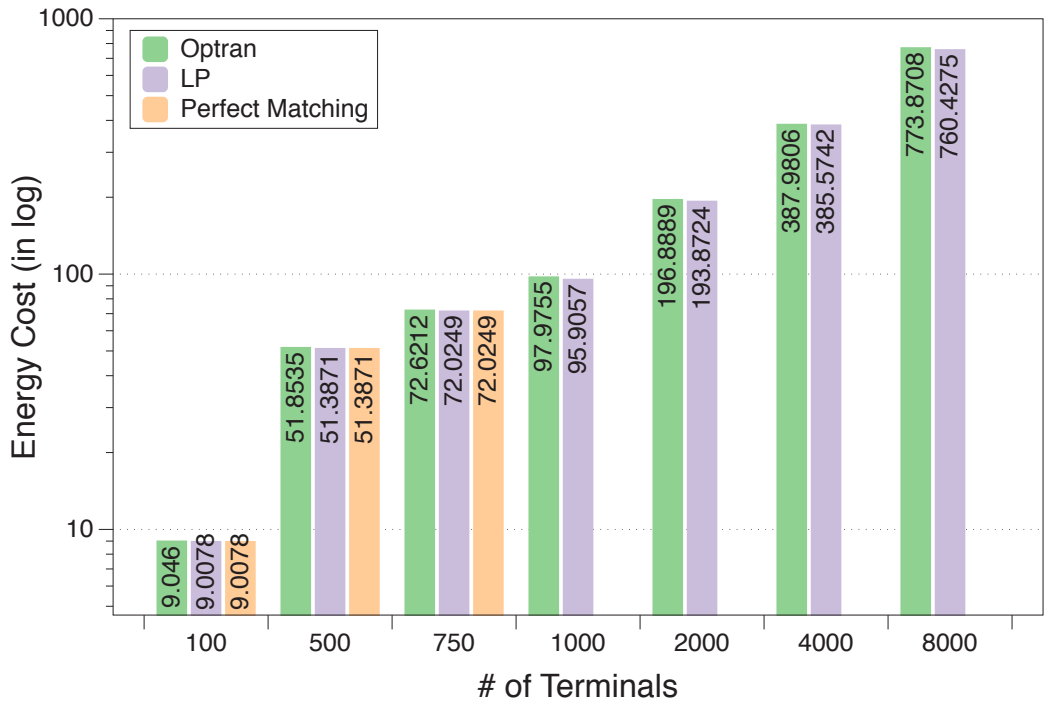

Fig. 3. A comparison of energy cost between Optran, LP, and perfect matching on different sizes of terminals. Number of base stations is fixed at 8, with equal capacity constraints.

2) *Computational Cost:* To discuss the difference of computation time and energy cost, we compare Optran, LP, and perfect matching solution with different number of terminals and base stations. Fig. 2 and fig. 3 show the comparison between terminals sizes differ from 100 to 8000, the number of base stations is fixed at 8 with equal capacity constraints, and the $\delta w$ is set to be $1e^{-4}$.

For energy cost perspective in fig. 3, all of the solutions performs well. In time perspective in fig. 2, Optran outperforms LP with all terminal sizes in magnitude order; in the case of 8000 terminals, the performance of Optran are up to 10000 times better than LP. The perfect matching yields the worst solution in computation time, and because of its cumbersome computational complexity, it only works on the cases that terminals sizes are smaller than 1000.

Notice that the computation time of LP and perfect matching grow exponentially while Optran grows linearly. This is mainly because that while computing the result of final clustering scheme, our method only requires $O(k)$ variables ($k$: base station) comparing to $O(nk)$ variables ($n$ terminals with $k$ base station) in LP. In fig. 4, the case of increasing base station number with 1000 terminals also supports this point.

In fig. 5, we demonstrate the ability of Optran on large scale of network. With 30000 terminals along with up to 2000 base stations in the domain, it clearly shows that the computation time of Optran grows linearly with the increase of the number of base stations.

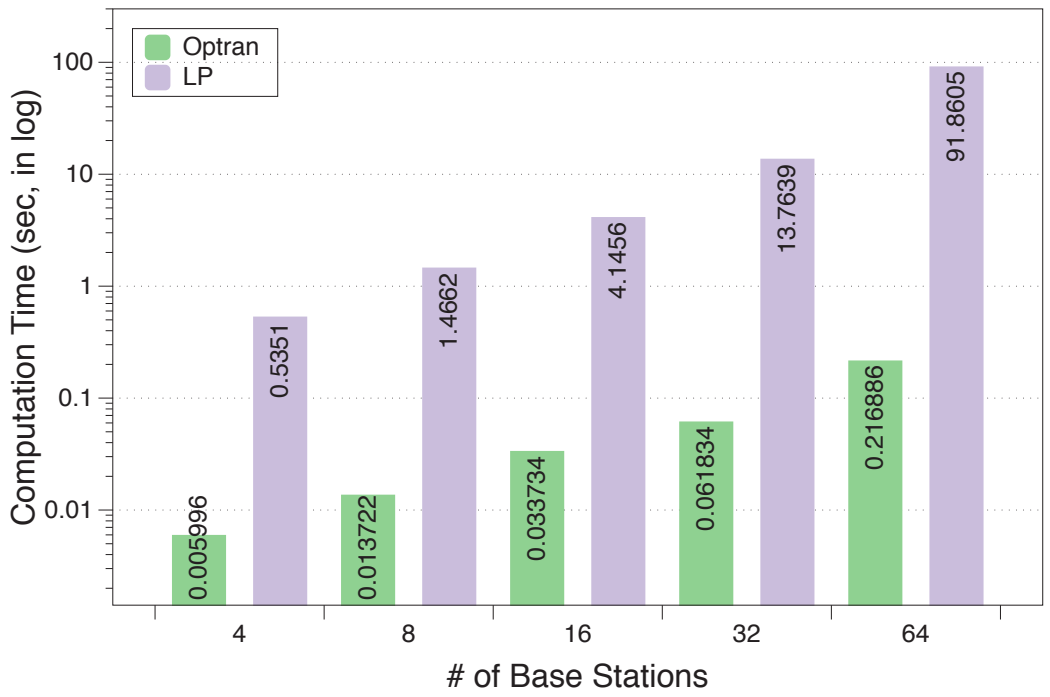

Fig. 4. A comparison of computation time of Optran and LP on 1000 terminals with different sizes of base stations with equal capacity constraints. The base stations are located randomly in domain.

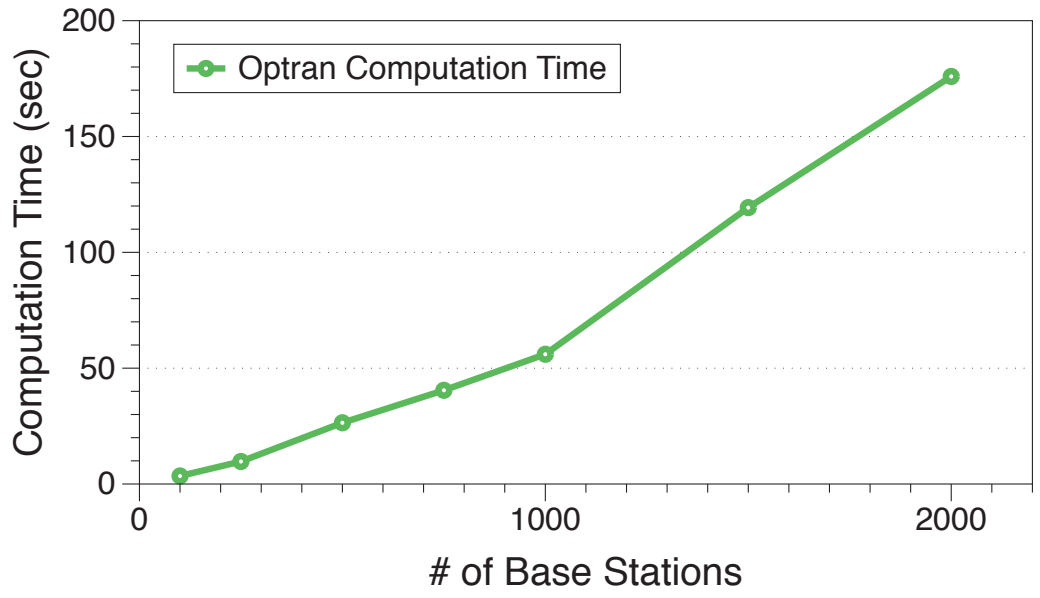

Fig. 5. Computation time of Optran growths linearly on 30000 terminals with up to 2000 of randomly distributed base stations with equal capacity constraints.

3) *Newton's Method vs Gradient Descend:* In Optran we used Newton's method, we compare with the gradient descend method in terms of computational cost. We refer the reader to fig. 6. With the same step size of 0.01 and over 4000 terminals, Optran uses a lot fewer iterations, sometimes two orders of magnitude better than gradient descend.

4) *Mobility:* To evaluate the performance of Optran under mobile setting, we first proposed a train moving scenario as fig. 9. In fig. 9, there are total 2150 users in the domain, each single user is represented by a "+" symbol. Suppose there is a train packed with users moving from left to right corresponding to time frame $t$; in order to get the kinetic clustering scheme for base stations, we compute the clustering for each snapshot. Since the power vector is similar between each snapshot (i.e. the Voronoi diagram is also similar), while we compute the next clustering scheme, we can re-apply the power vector (PV reapply in fig. 8) of the last snapshot as an initial guess for power vector. As stated in fig. 8, when we contiguously

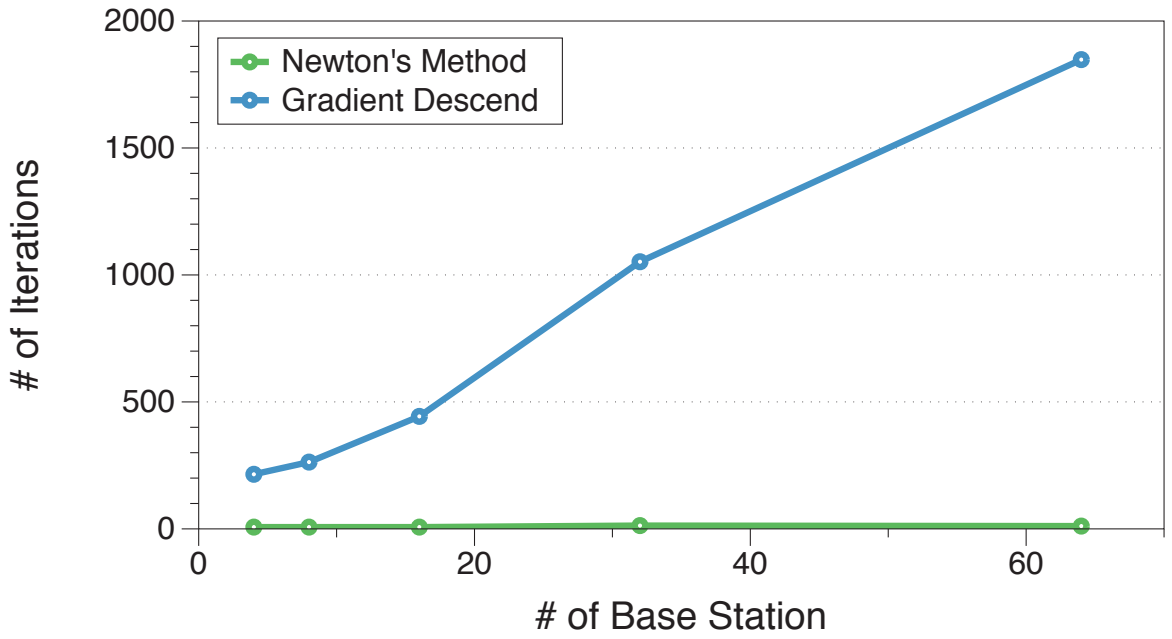

Fig. 6. Newton's Method vs Gradient Descend. With 4000 terminals and stepsize 0.01.

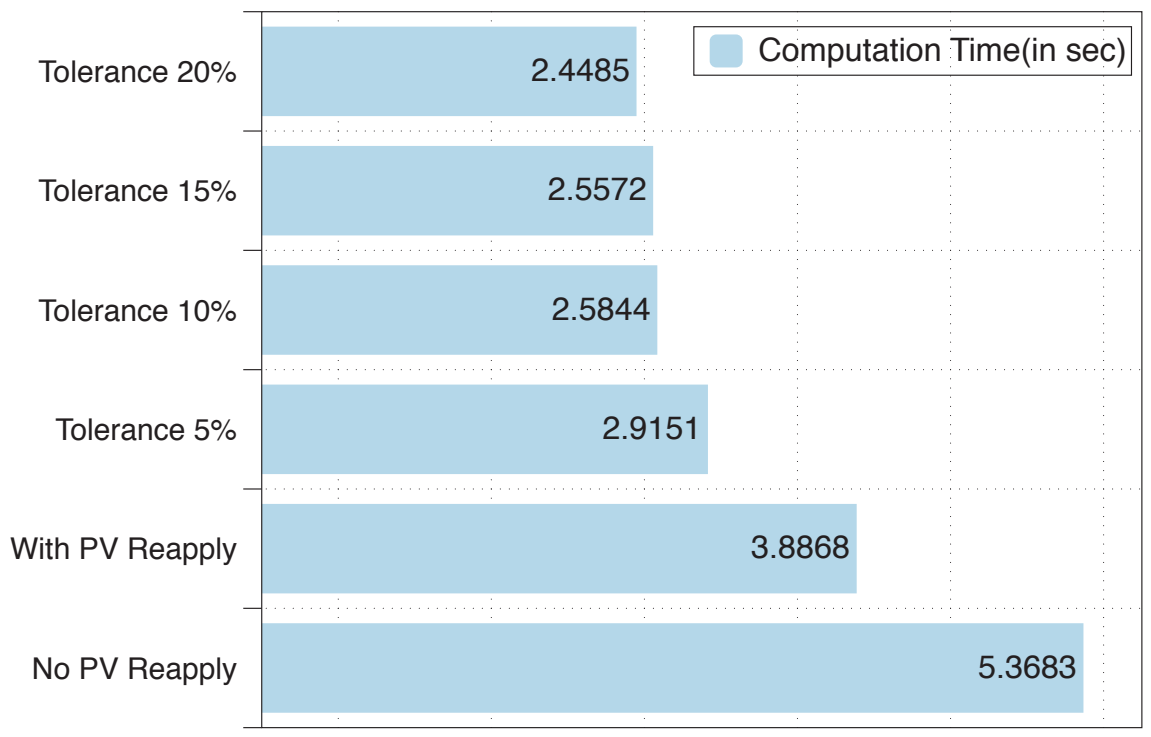

Fig. 7. A comparison of computation time with different capacity tolerance value in a linear motion setting with 3000 terminals and 8 base stations. The motion of the terminals are represented by 100 contiguous snapshots

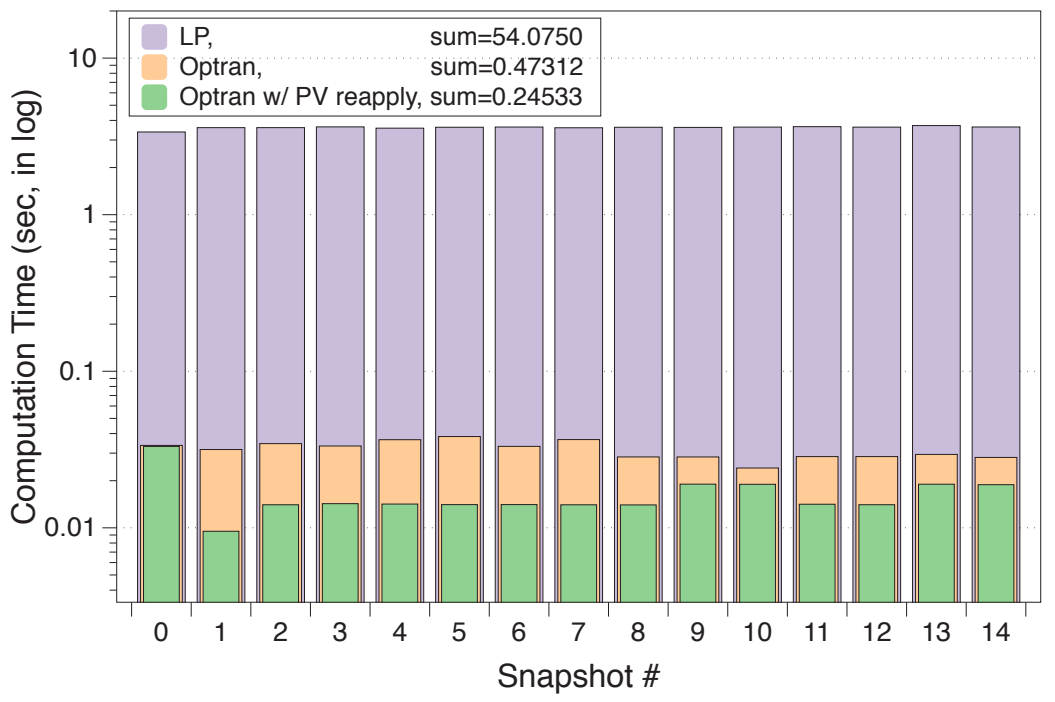

Fig. 8. A comparison of computation time between Optran and LP on a sequence of snapshot. Since in Optran the initial power vector can be inherent from the previous snapshot, the computation time is reduced after the first snapshot.

compute 15 snapshots, the time required drops 49%.

Last, we consider the trade off of computational time and the exactness of clustering. We consider the mobile setting that each user is on a linear motion: the terminal's mobility patten is $p + (q - p)t$ at time $t$ while $p$ and $q$ are randomly assigned. To further optimize computation, we assign the capacity tolerance to each base station, that is, we only re-cluster if one base station is out of capacity over the tolerance percentage. Fig. 7 illustrates the result with 100 contiguous snapshots. With proper tolerance as a tradeoff of computation time, the speed can be up to 30% better than pure PV reapply.

*5) Energy Consumption Model:* The proposed method is mainly for outdoor scenarios when energy consumption on wireless communication is proportional to the squared distance to the base station. For indoor situations, the energy consumption is proportional to $d$ to the power of $\alpha$, where $\alpha$ stays somewhere between 2 to 5, and $d$ is the distance to the receivers. Although the optimal transport theory did not cover this new optimization objectives, the algorithm works equally well in practice. In fact, the energy cost computed by our method is almost identical to results obtained by LP. The same level of speedup is observed.

## VI. Kinetic Clustering

In the discussion above we show a distributed algorithm for solving the kinetic clustering problem with good performance in practice. The second question we would like to understand is a fundamental one – how complex is it to compute the optimal solution for moving nodes for all time?

To formulate this problem rigorously, we take the approach of kinetic data structure [15], and assume that all nodes move continuously. However, by coherence of motion, the optimal solution only changes at discrete points of time. The idea of kinetic data structure is to track the discrete events and only update the solution when necessary. It does so by maintaining a set of *certificates* whose validity ensure the optimality of the solution. When any certificate fails, the structure is updated, and some certificates may be re-computed/re-evaluated. This ensures that the rate of updating the solution is adaptive to specific configuration and the object mobility patterns. Clearly the optimal solution can change arbitrary many times if the terminals/base stations follow arbitrary motion. So it is standard in kinetic data structure to assume that the motion trajectories of all terminals/base stations are pseudo-algebraic (e.g., specified by low-degree polynomials) such that any certificate of interest flips between true and false $O(1)$ times. Under such assumption we ask how many times the optimal solution would change. Below we provide upper and lower bounds for this value.

We first realize that for the static version, the optimal capacitated clustering has a unique solution but the solution can be realized by multiple power diagrams – due to the discrete formulation. In particular, consider a cell $C_j$ in the power diagram for a base station $y_j$. The set of points assigned to $y_j$ is the $Y_j$ which stay inside $C_j$. Now, one can (slightly) increase or decrease the weight $h_j$ while keeping the set $Y_j$ to be the same. The weight $h_j$ has a maximum value when a terminal in $Y_j$ stays on the boundary of $C_j$ and a minimum value when a terminal not in $Y_j$ stays on the boundary of $C_j$. Thus for any power diagram defining the optimal solution, we will change it so that any cell has at least one terminal on its cell boundary.

First, recall that if we uniformly increment/decrement the weights $h_1, h_2, \cdots, h_k$ by the same amount, this is equivalent to shifting the entire arrangement up and down. The structure of the upper envelope does not change. The projection of the upper envelope does not change. Thus we take one cell and fix its weight to be zero, without loss of generality, $h_1 = 0$ all the time. Now we take any cell $C_j$, we will increase $h_j$ (shrinking the cell $C_j$) until one terminal of $Y_j$ (say $x_i$) stays on the boundary. Let us suppose that $x_i$ stays on the boundary of $C_j$ and $C_\ell$. Now we say that $C_j$ and $C_\ell$ are in the same locked component. In particular, the height $h_j$ is taken such that $\text{Pow}(x_i, y_j) = \text{Pow}(x_i, y_\ell)$. Thus the difference $h_j - h_\ell$ is fixed by $||x_i - y_j||^2 - ||x_i - y_\ell||^2$. We denote this equation as a locking condition. We say $x_i$ is a locker.

Now we continue the same procedure. Take one locked component, Increase the weight of all cells in this component, which does not change the shared boundaries of the cells in this component, until a terminal stays on the outer boundary of this component. This will possibly grow a locked component to include one more cell, or merge two locked components into one. Eventually all cells belong to the same locked component and the cell boundaries with terminals on them correspond to a spanning tree $T$ of the dual of the power diagram. Since we set $h_1 = 0$ all the time. At the end of this procedure every cell is given a fixed weight. We say that the power diagram is at a locked configuration. Notice that first the combinatorial structure of this final power diagram may be different from that of the initial one; and depending on which cell to choose the above procedure may generate different final power diagrams. But all of these will define the same optimal clustering solution.

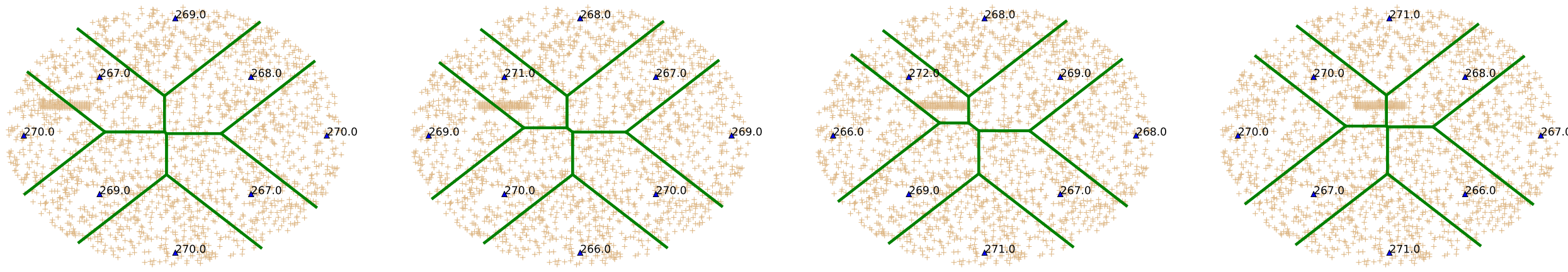

(a) $t = 0$ (b) $t = 2$ (c) $t = 4$ (d) $t = 6$

Fig. 9. A sample of terminal movement display by snapshots. In the scenario for mobile case, there is one train full of customers driving from left to right.

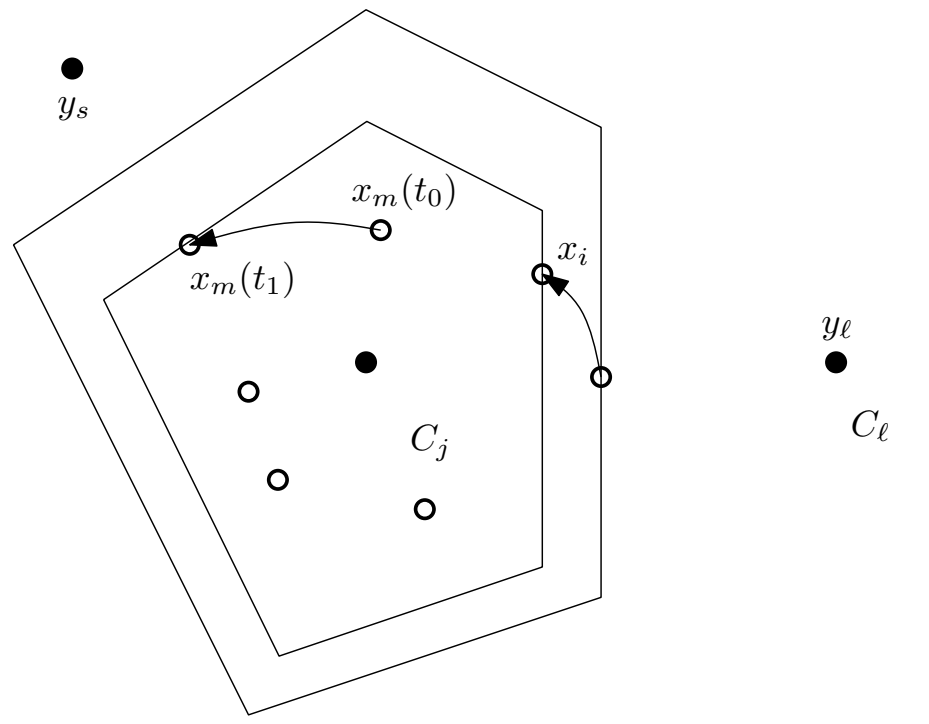

Fig. 10. Shrink cell $C_j$ by increasing $h_j$ until a terminal $x_i$ stays on the boundary.

The above discussion says that the number of weights for a base station that actually make a difference in the clustering solution is finite – if we reduce the power diagram by the above procedure. This is useful for the argument later.

Now we argue an upper bound on the number of changes to the kinetic clustering solution by using the power diagram formulation. Suppose all terminals and base stations start to move around and their positions at time $t$ are described by $x_i(t), y_j(t)$ respectively. And we have the initial power diagram such that base station $y_j$ has weight $h_j$ which is fixed. By the discussion above, we take the power diagram at a locked configuration defined by the spanning tree $T$ on the base stations. For a neighbor $y_j$ of $y_1$ on the tree $T$, we take its weight $h_j$ as a function of $t$, defined by $h_j(t) = ||x_i(t) - y_1(t)||^2 - ||x_i(t) - y_1(t)||^2$, where $x_i$ is the terminal on the boundary of power cells for $y_j$ and $y_1$. Similarly we propagate along the breadth-first search tree of $T$ starting from $y_1$ so that each weight is a function of $t$. The weights depend on the $k$ lockers and the tree $T$.

The above configuration will define a valid, optimal clustering solution until a critical event happens – a non-locker terminal $x_m$ hits a boundary of its power cell $C_j$ with a locker terminal $x_i$. See Fig. 10. When this happens, by the definition of the locker we know that

$$h_j(t_1) = h_\ell(t_1) + ||x_i(t_1) - y_j(t_1)||^2 - ||x_i(t_1) - y_\ell(t_1)||^2$$

and by the definition of the critical event we have that

$$h_j(t_1) = h_s(t_1) + ||x_m(t_1) - y_j(t_1)||^2 - ||x_m(t_1) - y_s(t_1)||^2$$

This gives us the equation:

$$\begin{aligned} & h_\ell(t_1) + ||x_i(t_1) - y_j(t_1)||^2 - ||x_i(t_1) - y_\ell(t_1)||^2 \\ = \; & h_s(t_1) + ||x_m(t_1) - y_j(t_1)||^2 - ||x_m(t_1) - y_s(t_1)||^2 \end{aligned}$$

Now, we can charge the number of critical events that lead to a change of the optimal clustering to such events. We only need to count how many such equations can be true if the base stations and terminals follow pseudo-algebraic motion. In each equation three base stations and two terminals (one of them is a locker) are involved. Further, we also have the weight of the two base stations whose functions are defined by possibly $k-1$ locker terminals and the tree $T$ that they define. This give us a total number of changes by $n \cdot \binom{n}{k-1} \cdot g(k-1)$, where $g(k)$ is a function of $k$. If we assume that $k$ is a constant, the above bound is $O(n^k)$.

## VII. Related Work

We survey related work in two directions, prior work on algorithms for optimal transport problem and prior work on the application of capacitated clustering in computer networking.

The optimal transport problem has been extensively studied in the mathematics community. Almost all the study there focus on the existence, uniqueness, structural properties of the solution and its connection to other theories. This has been surveyed earlier. On the algorithmic side, most work is on the discrete formulation, as in our case. Beside the min cost matching or LP formulation, the most notable algorithm is the one in Aurenhammer et al. [4]. The authors suggested two algorithms, the first one is a combinatorial algorithm utilizing the power diagram formulation with running time $O(k^2 n \log n + kn \log^2 k)$ for $n$ terminals and $k$ base stations. The second algorithm is a gradient-descent method to find the optimal solution, different from the one used here. They worked on the static setting only. In [16], the authors start with an arbitrary partition that fulfills the capacity constraint without representing a valid Voronoi diagram, then iteratively swap the assignment of points to sites guided by an energy minimization. This is easy to implement but the convergence was not proved. Last, a number of other heuristic algorithms were developed [17] with no theoretical guarantee.

On the other hand, in applications of optimal transport theory and capacitated clustering, all of them use Linear Programming solvers. We discuss these applications below.

Because of its resource allocation aspect, optimal transportation theory is mostly applied for load balancing and resource management in computer networking. In [18], it is utilized to sooth the congestion of the network. When a network is under congestion, probabilities as costs are assigned to each used route, and Monge-Kantorovich approach is obtained for finding a best allocation for packages. In [19], optimal transportation problem is applied to solve the resource allocation problem for queueing process while considering the extra congestion cost for each user in queue. In [20], the authors applied optimal transportation problem to solve the base station assignment problem under congestions. In their work, the LP based Monge-Kantarovich approach is applied for assigning users to base station while minimizing the transmission cost.

In wireless networks, the problem of optimizing the resource allocation between base stations and terminal users has been extensively studied. In [21], the authors proposed a multiuser sub-carrier allocation problem over OFDM system. This problem describes the minimization problem of the down link transmission power from the sub-carrier to users while fulfilling the user's data transmission requirement. The problem is modeled as a bipartite matching problem, the Hungarian algorithm [14] is suggested for the optimal allocation result. More follow up sub-carrier allocation research based on linear programming (LP) solution can be found in [22]–[24].

In [25], Yates et al. proposed an iterative algorithm based on LP to solve Minimum Transmitted Power (MTP) problem on cellular network. The MTP problem concentrates on finding an allocation between user and base station such that minimizing the total transmitted uplink power, and also maintain the required carrier to interference ratio (CIR) of each user. In order to simplify, this MTP problem can be remodeled as the uplink assignment problem with minimal cost. However, the capacity of the base station is not considered in this paper, a base station located in a dense user area might be assigned for users whose number exceeds its capacity. As a result, the overall performance is reduced. Moreover, since the solution is based on LP, the space requirement of the assignment vector is based on the size of users, which can be huge and lead to high computation cost in large scale cases.

## VIII. Conclusion

This paper initiated the study of kinetic capacitated clustering problem and provides an algorithm that substantially improve the state of the art. Our algorithm is based on the optimal transportation problem that recently realized to be valuable for resource allocation like problem. We showed that for our method, the number of variables used is only $O(k)$ compared with $O(nk)$ in LP solutions, where $k$ is the number of base stations and $n$ is the number of mobile nodes. In future work we would like to 1) find a tighter bound on the number of changes of capacitated clustering problem in the kinetic data structure framework; 2) apply the algorithms in practical applications.

## Acknowledgment

The authors are partially supported by grants from AFOSR (FA9550-14-1-0193) and NSF (CCF-1535900, DMS-1418255, DMS-1221339, CNS-1217823).